\documentclass[runningheads]{llncs}
\usepackage{amsmath}
\usepackage[T1]{fontenc}
\usepackage{graphicx}
\usepackage{xcolor}
\usepackage{amssymb}
\usepackage{bm}
\usepackage{subcaption}
\usepackage{booktabs}

\newcommand{\change}[1]{{#1}}
\newcommand{\remove}[1]{}

\newcommand{\rec}{\bm{r}}
\newcommand{\send}{\bm{s}}
\newcommand{\signal}{x}

\usepackage{pgfplots}
\DeclareUnicodeCharacter{2212}{−}
\usepgfplotslibrary{groupplots,dateplot}
\usetikzlibrary{patterns,shapes.arrows}
\pgfplotsset{compat=newest}

\definecolor{sonnet_red}{RGB}{214,39,40}
\definecolor{gccphat}{RGB}{255,127,14}
\definecolor{groundtruth}{RGB}{44,160,44}
\definecolor{sonnet}{RGB}{31,119,180}
\definecolor{sonnet_aqua}{RGB}{102,205,170}
\definecolor{sonnet_magenta}{RGB}{255,0,255}

\begin{document}
\title{SONNET: Enhancing Time Delay Estimation by Leveraging Simulated Audio\thanks{This work was partially supported by the strategic research project ELLIIT and partially supported by the Wallenberg AI, Autonomous Systems and Software Program (WASP) funded by the Knut and Alice Wallenberg Foundation. Computational resources were provided by the Swedish National Infrastructure for Computing at C3SE and NSC, partially funded by the Swedish Research Council, grant agreement no. 2018- 05973.}}
\titlerunning{SONNET: Enhancing Time Delay Estimation}
% If the paper title is too long for the running head, you can set
% an abbreviated paper title here
%
\author{Erik Tegler\orcidID{0000-0002-8730-8301} \and
Magnus Oskarsson\orcidID{0000-0002-1789-8094} \and
Kalle Åström\orcidID{0000-0002-8689-7810}}
\authorrunning{Tegler et al.}

\institute{Lund University, Lund, Sweden \\
\email{\{erik.tegler, magnus.oskarsson, karl.astrom\}@math.lth.se}\\}
\maketitle              % typeset the header of the contribution
\begin{abstract}
Time delay estimation or Time-Difference-Of-Arrival estimates is a critical component for multiple localization applications such as multilateration, direction of arrival, and self-calibration. The task is to estimate the time difference between a signal arriving at two different sensors. For the audio sensor modality, most current systems are based on classical methods such as the Generalized Cross-Correlation Phase Transform (GCC-PHAT) method. In this paper we demonstrate that learning based methods can--- \change{even based on synthetic data}---significantly outperform GCC-PHAT on novel real world data. To overcome the lack of data with ground truth for the task, we train our model on a simulated dataset which is sufficiently large and varied, and that captures the relevant characteristics of the real world problem. We provide our trained model, SONNET (Simulation Optimized Neural Network Estimator of Timeshifts), which is runnable in real-time and works on novel data out of the box for many real data applications\change{, i.e.} without re-training. We further demonstrate greatly improved performance on the downstream task of self-calibration when using our model compared to classical methods.
%Time-difference-of-arrival estimates of audio and radio signal are crucial for a number of applications such as localization, mapping and receiver-sender position calibration, which in turn are important, e.g.\ for microphone array calibration, speaker diarization, beam-forming. Most current systems are based on cross-corelation techniques and variants thereof, for example the GCC-PHAT method. In this paper we demonstrate that using both a sufficiently large and varied simulated dataset and a sufficiently rich deep learning network architecture, it is possible to improve significantly on current state-of-the-art and provide a method for time-difference-of-arrival estimate that works out of the box for many real data applications without re-training. 
\end{abstract}

% Sections ---------------------------------------------------

% Introductionm

\section{Introduction}

\begin{figure}
    \centering
    \includegraphics[width=\textwidth]{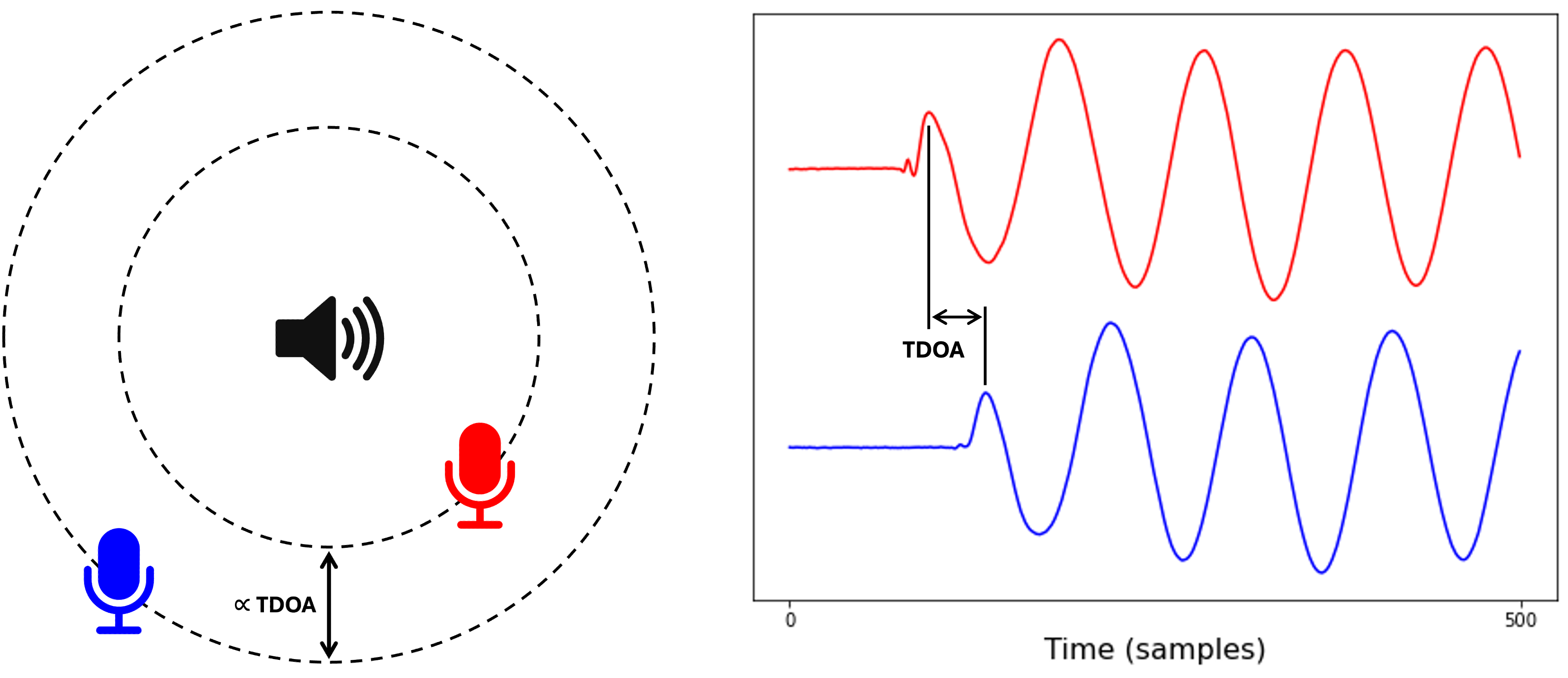}
    \caption{Since the microphones are at different distances from the speaker, the signal arrives at different times for each of them. By estimating the timeshift in the signals (right figure), and combining it with the propagation speed of the signal, we get a measurement of distance difference (left figure).}
    \label{fig:TDOA}
\end{figure}

% 1. What is the problem?
Time Delay Estimation (TDE) is the problem of determining how much later (or earlier) a signal from a transmitter is received at two different receivers. The result is often denoted Time-Difference-Of-Arrival (TDOA), see Fig~\ref{fig:TDOA}.
% 2. Why is it interesting and important? 
%Time Delay Estimation (TDE) 
TDE is a pivotal problem, primarily due to its critical role in localization and positioning systems. By enabling the determination of the TDOA of a signal at different receivers, TDE provides the foundational measurements for inferring the spatial location of senders and/or receivers using further methods. Examples of such methods are:
\begin{itemize}
    \item Multilateration, where the positions of the receivers are known and the TDOA \change{estimates} are used to estimate the position of the sender, \cite{GustafssonRT03_11,larsson_optimal_2019}.
    \item Direction of arrival, where prior knowledge of the geometry of a receiver array is used together with TDOA measurements from multiple pairs of receiver to compute from which direction the signal is received. 
    \item Self-calibration, where the positions of both receivers and senders are \change{estimated} solely based on the measured TDOA, \cite{zhayida2014automatic,larsson2021fast,tegler_eusipco_2022}. 
\end{itemize}

%Through the analysis of both radio and audio signals it is possible to extract meaningful geometric estimation. One such example is localization, \cite{GustafssonRT03_11,larsson_optimal_2019} i.e.\ determining the position of for example the receiver using known transmitter positions. Another example is the simultaneous determination of both transmitter and receiver positions, \cite{zhayida2014automatic,larsson2021fast,tegler_eusipco_2022}. 

Accurate localization of sender and receiver nodes is crucial for various applications, including microphone array calibration, speaker diarization, beam-forming, radio antenna array calibration, mapping, and positioning \cite{plinge2016acoustic}. 

In these and many other applications, the initial signal processing step of obtaining reliable TDOA estimates, plays an important role. Currently, the state-of-the-art method is the Generalized Cross-Correlation Phase Transform (GCC-PHAT) and its variants \cite{Knapp1976}. However, recent research indicates that there is substantial room for improvement. In \cite{astrom2021extenstion} it was shown that the average performance of existing methods fell below the desired threshold in nearly 40\% of estimations based on a real dataset with ground truth.  

The TDE problem is relevant across multiple different signal modalities such as audio and radio. However, in this paper the main focus is directed towards the analysis of audio, although 
\change{studying radio signals is an interesting subject for future studies.}
%\change{studying radio signals is an interesting further application.}
% it would be interesting to see how the methods could be applied to radio signals. 

% Time delay estimation (TDE) stands as a fundamental challenge in signal processing, central to a wide array of applications ranging from 
%mapping \cite{tegler_eusipco_2022,} and 
%localization %\cite{GustafssonRT03_11,larsson_optimal_2019} to 
% \cite{Meldercreutz1741,ChampagneBS96_4, DiBiaseSB01, GustafssonRT03_11, KidronSE07_55, GilletteS08_15, HoS08_56, AlamedaH14_22}
% beamforming
% diarization?
% tracking
% robotics

%\todo{...more applications: maybe tracking robotics from Axels paper}. 

\subsection{Challenges}
% 3. Why is it hard? (E.g., why do naive approaches fail?)
TDE in the audio domain presents a set of challenges that significantly complicates the estimation process. Unlike scenarios involving controlled signal transmission, for a variety of audio applications the sound source is not a controlled entity. Therefore our estimation techniques need to handle unknown signals with unpredictable characteristics. This difficulty is compounded by reverberations, a common phenomenon in acoustic spaces where sound waves reflect off surfaces, creating multiple delayed echoes that can obscure the true signal path. Another challenging aspect arises when dealing with moving sound sources, as this introduces a dynamic element to the TDE problem  \cite{flood2019stochastic}. As the source moves, the relative distances to the receivers change continuously, altering the TDOA in real-time and demanding adaptive estimation techniques capable of handling these variations. Together, these factors—unknown signal characteristics, reverberations, and source mobility—make audio-based TDE a particularly demanding task, necessitating sophisticated algorithms and approaches to achieve reliable estimation.

% 4. Why hasn't it been solved before? (Or, what's wrong with previous proposed solutions? How does mine differ?)
\subsection{Related works}

%\todo{Unclear if I should have a longer Related works section with specifics such as: Learning in Audio in general, simulated data for learning in general, etc.}

Previous methods for TDE in audio signal processing exhibit notable limitations in handling complex real-world scenarios. The Generalized Cross-Correlation method, and specifically its Phase Transform variant GCC-PHAT is often used as a starting point. It is robust to measurement noise and works well across diverse signal types. However, GCC-PHAT shows limitations when dealing with reverberations and moving sound sources. 

%\todo{unclear how to say this when I don't know exactly why it doesn't work. But theoretically it should not work while we have reverberations and moving sound source}

In \cite{flood2019stochastic} methods were developed for estimation of TDOA with sound source or receiver motions. These methods were based on local optimization of initial estimates based on the GCC-PHAT. 

Recent advancements have seen the adoption of machine learning techniques for TDE. For small baseline receiver arrays, where the receivers are typically placed equidistant, direction of arrival can be made using alternative techniques, for example steered-response power with phase transform \cite{diaz2020robust}, spectrograms \cite{wang2023fn} and raw waveforms \cite{he2022sounddoa}. For DOA estimation both traditional and data driven methods have been used successfully. 

For large baseline arrays, where the receivers are placed in an ad-hoc fashion there has been some attempts at data-driven methods, see for example \cite{vera2018towards,grinstein2023dual,grinstein2023graph,gong2022end,feng2023soft} and \cite{grumiaux2022survey} for an extensive overview. Many of these are, however, trained and evaluated for specific subtasks. One of the problems has been the lack of real world data with accurate ground truth. Previous work circumvents this by using simulated data to train their models. While this approach is promising, these papers make no claim to have a model which works on novel real world data. Instead, only training and evaluating on similar simulated datasets. 

% 5. What are the key components of my approach and results? Also include any specific limitations.

\subsection{Contribution} 
In this paper we propose more careful modeling of the problem, and thereby improving  over previous learning methods by increasing the scope and quality of the simulations. Many learning based approaches focus primarily on reverberation effects, neglecting other critical factors. Thus, limiting the generalizability of these methods to real data.  This paper aims to contribute to the body of knowledge on TDE by leveraging simulated data to explore and enhance estimation techniques for audio signals. To summarize, our \change{contributions are}:
\begin{itemize}
    \item Demonstrating that data driven models trained on large scale simulated sound \change{datasets}, generalize to real data as well as to novel sounds for the TDE task.
    \item Providing a model, SONNET - Simulation Optimized Neural Network Estimator of Timeshifts, which outperforms state of the art methods for TDE and is evaluated on both simulated and real data.\footnote{Code available at: \url{https://vision.maths.lth.se/sonnet/}}
    \item Demonstrating how the new estimators improve performance on downstream tasks.
\end{itemize}

\section{Problem setup}

% techincal describtion of problem setup
Consider a reverberant room containing two receivers positioned at $\rec_1,\rec_2 \in \mathbb{R}^3$ and a moving sender located at $\send(t) \in \mathbb{R}^3$. %\todo{define t or not?}. 
The sender is emitting an unknown signal $\signal (t)$ which is being recorded by the receiver at $\rec_i$ as the signal $x_i(t)$.
The TDOA at time $t$ for receivers $i$ and $j$, is defined as
\begin{equation}
    \Delta (t) = \frac{||\rec_i - \send(t)|| - ||\rec_j - \send(t)||}{v_x},
\end{equation}
where $v_x$ is the propagation speed of the signal. We are in this paper primarily interested in this direct path TDOA, but an interesting extension would be to also consider TDOA measurements corresponding to multi-path components from reflective planes, that could potentially provide richer information, \cite{dokmanic2014localize,zhayida2016automatic}. 

%We will show that instead of constructing an estimator by hand, e.g. GCC-PHAT, we can construct an estimator using machine learning techniques on generated data.

For the modeling we assume that the received signal  $x_i(t)$ can be modeled as
\begin{equation}
    x_i(t) = \int h_i(t - \tau,\tau)\signal(\tau)d\tau + \epsilon_i(t),
\end{equation}
where $\epsilon$ is the noise and $h_i(t, \tau)$ is the impulse response from the sender to receiver $i$ at the position $\send (\tau)$. The impulse response captures the acoustic properties (position, orientation) of both the receiver and sender as well as the reverberant properties of the room. Typically there is a strong direct path component in the impulse response, corresponding to a time-delay of $\frac{||\rec_i - \send||}{v_x}$, which allows for the TDOA estimation. While the TDOA is time dependent, we will for the rest of this paper refer to the TDOA of a pair recorded signals $\signal_i(t), \signal_j(t), 0 < t < T$ as the TDOA value at the middle of the signal, i.e. $\Delta (\frac{T}{2})$. The goal is to use two recorded signals $\signal_i, \signal_j$ to estimate this TDOA value. 

%\todo{\begin{itemize} 
%\item Definiera diskret?
%\item Olika beteckningar på $x$ och $x_i$?
%\item Brusmodell?
%\item Begränsningar modell?
%\end{itemize}}

% describe similarities to work done by Berg et al. Similarities are: regression-via-classification, pyroomacoustics gives reverberations
%\subsection{Simulation}

% Goal of section: 
% Explain 

\section{Data simulation}
Similar to earlier work by Berg et al \cite{berg22_interspeech}, we use Pyroomacoustics \cite{scheibler2018pyroomacoustics} to compute impulse responses using the image source method \cite{allen1979image}. However, we augment the simulation in a number of important ways. Instead of simulating a single room we simulate a broader class of rooms in order to cover a larger set of possible impulse responses. Also, with the goal of making the simulation better reflect reality, we both simulate a moving sound source and also microphones and sound sources which are not omnidirectional. How and why will be explained in more detail in the following sections and motivated by our ablation study in section~\ref{exp:abl}.

%contrast to prior work, we significantly expanding the volume of the simulated dataset as well as enhance the level of detail within the simulation to more closely resemble real-world data. Two significant enhancements in our simulations, compared to earlier efforts, are the inclusion of moving senders in the simulation and the use of directional microphones. These enhancements are driven by the goal of creating simulations that more accurately reflect real-world conditions.  This is further supported by the results in the ablation study, see section~\ref{exp:abl}. 

% Moving sound source
\subsection{Moving sound source}
To simulate a moving sound source we first generate a path by constructing a quadratic Bézier curve $s(t), \, 0 < t < T$ with a length shorter than some maximum length. Because simulating the sound from a moving sound source is difficult within the Pyroomacoustics framework, we instead discretize the curve into $k$ points 
\begin{equation}
    \lbrace s(t_1), \dots , s(t_k) \rbrace, \quad t_i = \frac{i-1}{k-1}T.
\end{equation}
The sound from a moving sound source is then approximated by dividing the played sound $x(t)$ into $k$ equally sized parts and simulating part $x(t), \frac{i-1}{k}T < t < \frac{i}{k}T$ as a stationary speaker at point $s(t_i)$. This is essentially simulating that the sound source is jumping to a new location along a path after each time $\frac{T}{k}$. Following this methodology, the signal is computed as a sum of convolutions
\begin{equation}
    \signal_i(t) = \sum_{j=1}^{k} h_i(t, \frac{j}{k}) * \bar{x}^{(j)}(t),
\end{equation}
where $\bar{x}^{(j)}$ is the $j$th part of the signal, zero-padded to have the same shape as the original signal $\signal$. 
%We simulate a moving sound source as follows: The sound file is divided into \(d\) equally long segments. Additionally, an intended path for the moving source is outlined. To simulate the sound source traversing this path, a segment of sound lasting \(T/d\) seconds is played from a specific location along the path. Subsequently, the next \(T/d\) second sound segment is played from the following location on the path. By sequentially playing sound clips from discrete points along the intended trajectory, we effectively simulate a sound source that 'teleports' a short distance along the path every \(T/d\) seconds. 
%This approach is utilized because simulating a continuously moving sound source introduces increased complexity. 

%Although the TDOA is well defined for each time instant $t$, because of the time delay to the receivers it is not obvious what time interval at the receivers to use for this estimation. These errors are however small in the intended applications, since the speed of the sound source is small compared to the speed of sound.

%\todo{Also need to talk about how the ground truth is not well defined for a time-window, we elect to just choose the TDOA value at the middle of the interval.}

% Directionality
\subsection{Directionality}
Previous work simulated both 
\change{microphones and speakers} 
%microphone and speaker 
as being omnidirectional, which means they emit/receive their signal equally well in all directions. However, the directional dependence in reality is rather complicated, since it depends on what hardware is used. This is further complicated by the directionality of a microphone not being constant for all frequencies, as demonstrated in \cite{zetterqvist2023using}. We settled on using the subcardioid sensitivity pattern, since it is a common model for directional microphones and therefore already implemented in Pyroomacoustics \change{\cite{de2011generalized}}.

% Details on how to generate and store data efficently

%\subsection{Storing dataset}
%Each training example consists a tuple 
%\begin{equation}
%    (\signal_1, \signal_2, \Delta T). 
%\end{equation}
%To minimize computation and storage requirements, rather than simulating two microphones for each signal, we simulate and store $N$ microphones. We then save the tuples
% \begin{equation}
%     (x_i, \frac{1}{v_x}||\rec_i - \send||)
% \end{equation}
% as our dataset. The pairs can then be constructed at training time by combining two tuples corresponding to the same simulation. In this way we generate and store $\binom{N}{2}$ examples for approximately $N$ times the computational and memory cost. The drawback is reduced diversity of examples in the dataset.

% sound noise? & ground truth noise? for regularization

% Exact description of how dataset was simulated, i.e. list of constants etc.

\section{Inference model}
\begin{figure}
    \centering
    \includegraphics[width=\linewidth]{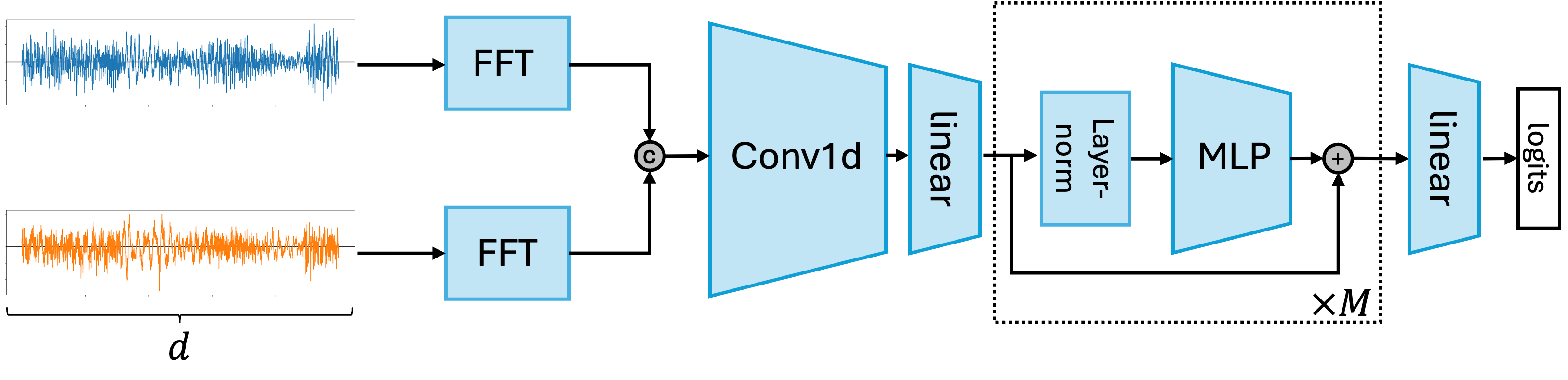}
    \caption{\textbf{System overview:} Our model takes two audio recordings of length $d$ as input data. The data is first converted to the frequency domain, using the fast Fourier transform, and stored with real and imaginary components as different channels. It is then sent through a series of 1d convolutional layers. The features are then processed using $M$ stacked pairs of linear layers along with skip connections. Finally, the logits are acquired by adding a linear layer after the last residual block.}
    \label{fig:net_arch}
\end{figure}

One of the main contributions of this paper is to show that it is possible to close the sim2real gap, i.e.\ to demonstrate that simulated data generalizes to real data, given enough simulations of sufficiently rich character. Therefore we opted for a network architecture with the properties of simplicity and being easily trainable, see Fig \ref{fig:net_arch}. The main part of our network is a ResNet, since it is a simple architecture which is easily trainable. However, an issue with audio data is that it has low information density, making it difficult to input the audio data directly into the ResNet, without significantly increasing the number of parameters of the model. We therefore use two common approaches to compress the audio data.

First we use a fast Fourier transform and only store the values for frequencies which are below some threshold frequency $f_{max}$. Another advantage of using the Fourier transform is that it mitigates the problem of the spectral bias in neural networks \cite{rahaman2019spectral}. Secondly, we use a backbone of 1d convolutional layers to extract more dense features from the data. 

In the same manner as \cite{berg22_interspeech} our model performs regression-via-classification (RvC). This means that our network does not output TDOA values, but rather outputs logits for a fixed number of classes. Each of the classes then corresponds to a range of TDOA values.

To make the data have the right size we use linear layers as projections between the backbone and the ResNet, as well as a linear classifier between the ResNet and the final logits.

%\todo{Det är något med språket här, som jag funderar på 'We considered' ...}
%We considered opting for the architecture suggested by Berg et al, as it likely introduces model biases which are helpful for the TDE task. However, the architecture also has disadvantaged such as being more complicated as well as having significantly longer inference time compared to GCC-PHAT, which is why we chose a simpler architecture.

\section{Implementation details}
%\subsection{Simulation \& Training details}
In our code we have provided a pretrained model SONNET. In this section we have outlined implementation details on both how SONNET was trained, as well as how the dataset it was trained on was generated.

The sound used when training the model are from the \textit{Musan} dataset \cite{musan2015}, which is an openly available corpus containing about 18 GB of music, speech and noise. For our simulations we used a signal length of $d$=10,000 samples with a sampling frequency of 16 kHz. However, since we want reverberations to be present at the beginning of the recording we simulated 2000 extra samples at the beginning of each simulation. For each recording, we simulated a new room in the shape of a rectangular cuboid with each of the three dimensions having a length uniformly sampled in the interval $[1,10]$ m. The reverberation level in the room was varied by sampling the reflection coefficient of the walls from the interval $[0.05,0.99]$. The path of the sender was simulated in one of two different ways with equal probability, either as a stationary point source or as a randomly sampled quadratic Bézier curve with maximum velocity of 5 m/s. In each room we simulated 50 microphones recording the signal.

The dataset consist of 10,000 rooms, which means that it in total contains $10,000 \binom{50}{2} = 12$ million training examples of pairs of recordings. The memory footprint of the dataset is 19 GB.

For the model we set our threshold frequency $f_{max} = 4800$ Hz. The backbone consisted of three 1d convolutional layers. The ResNet consisted of $M=4$ blocks. Throughout the network we used GELU \cite{hendrycks2016gaussian} as activation function. We choose to have 1000 output classes as possible predictions for our model. Each class corresponded to bins of TDOA values with a width of 1 sample. This means our model makes predictions with the same resolution as GCC-PHAT.

The model was trained in PyTorch \cite{Paszke_PyTorch_An_Imperative_2019} with the AdamW optimizer \cite{loshchilov2017decoupled}, a batch size of 4096, a learning rate of 0.0003, during 20 epochs. We used the cross-entropy loss with label-smoothing of 0.1 as our loss function.
The training was done on Tesla V100-PCIE-16GB GPU and took 3 hours.

% \begin{table}[]
% \caption{Simulation details}
%     \centering
% \begin{tabular}{cc}
%     \toprule
%              Constant, $d$ & Value/Range of values \\ \midrule
%          signal length, $d$ (samples) & 10,000 \\ 
%          Reverberation constants & 0.15 \\
%          Maximum speed (m/s) & 5\\
%          SNR & \\
%          ... & \\ 
%     \end{tabular}
    
%     \label{meth:sim_details}
% \end{table}

%We train two models \modelnamemusic on only the music part of the dataset, and \modelname on the entire dataset. The reason we train the model on only one class of sound is so we can evaluate it on the other classes. This will give us an idea of how well the model generalizes to new classes of sound. More simulation details are provided in Tab \ref{meth:sim_details}. \todo{Complete description of simulation in Appendix?}

%\subsection{Training details}

% consider picking a network from previous work. My hypothesis is that it will not matter that much as long as the network is not way to small. If the network used by others is way to small, then I have a point of comparisson at least

% experiments
\section{Experiments}
% viable path to out-of-the-box detector, not claiming we are using the best architecture.  
% Budskap angående jämförelse med andra metoder: Flera andra metoder använder liknande arkitektur som Axel dvs, maxiumum of 23 samples difference i.e. ~0.5 meters. detta gör jämförelser med deras metod svår eftersom de inte kan skatta skillnader större än 0.5 meters, att begränsa sig till problem som bara har 0.5 meters skillnad skulle också vara konstigt då att gissa rätt bland 10/47 val är betydligt lättare än 10/1000 val (10 eftersom det finns 10 gissningar som är inom 10 cm från svaret).

In this section we first analyze the inference speed and memory footprint of the proposed system. We then study the performance of the system on both simulated and real data, \change{by making several comparative studies against GCC-PHAT}. Finally we show how the proposed system can be used to improve a previous state-of-the-art system for automatic self-calibration of an ad-hoc configuration of microphones. 
We quantitatively and qualitatively (see Figures~\ref{fig:sensitivity}, \ref{fig:acc_at} and \ref{fig:qual_res}) show that our model outperforms previous methods for performing TDE in novel real world settings. 
%While this paper builds upon work done by \todo{Axel and of papers refering to him}, neither of these methods makes any claim to be useful out-of-the-box for novel real world scenarios. Both papers have not evaluated their method on real world data and they are also making design choices which limits the general applicability of their methods such as having a maximum detectable TDE of 23 samples $\approx$ 0.5 m.    
%comparing results with these methods would be an apple

\subsection{Inference speed and Memory Footprint}
The ideas in this paper can be used train TDE models of different sizes, and can therefore be tailored to the available memory and computation requirements for a specific use case. However, we suggest as a starting point to use SONNET, which we have provided along with this paper. SONNET has 20 million parameters and a memory footprint of 75 MB. 

Inference speed was evaluated on both CPU, Intel(R) Xeon(R) W-2125 CPU @ 4.00GHz, and on a GPU, Tesla V100-PCIE-16GB. The results are shown in Table \ref{tab:inference_speed}. To summarize, SONNET takes four times longer to run than GCC-PHAT, however, the inference time is still fast enough to run SONNET on a CPU in real-time without any issues.

\begin{table}[]
    \centering
     \caption{Computation time per pair using a batch size of 100 recording pairs.}
     \vspace{10pt}
    \label{tab:inference_speed}
    \begin{tabular}{lcc}
    \toprule
%        \hline
         & SONNET (ms) & GCC-PHAT (ms) \\
         \midrule
    %    \hline
        CPU & 0.94 & 0.32 \\  
        GPU & 0.022 & 0.005 \\
        \bottomrule
    \end{tabular}
   
\end{table}

\subsection{Noise and reverberation sensitivity (simulated data)}
\begin{figure}
    \centering
    \begin{subfigure}{0.49\textwidth}   
        \centering
        \scalebox{0.72}{% This file was created with tikzplotlib v0.10.1.
\begin{tikzpicture}

\definecolor{darkgray176}{RGB}{176,176,176}
\definecolor{darkorange25512714}{RGB}{255,127,14}
\definecolor{lightgray204}{RGB}{204,204,204}
\definecolor{steelblue31119180}{RGB}{31,119,180}

\begin{axis}[
legend cell align={left},
legend style={
  fill opacity=0.8,
  draw opacity=1,
  text opacity=1,
  at={(0.03,0.97)},
  anchor=north west,
  draw=lightgray204
},
tick align=outside,
tick pos=left,
x grid style={darkgray176},
xlabel={SNR (dB)},
xmin=-33, xmax=33,
xtick style={color=black},
y grid style={darkgray176},
ylabel={Inlier ratio @ 10cm},
ymin=0, ymax=1,
ytick style={color=black}
]
\addplot [very thick, steelblue31119180]
table {%
-30 0.281598513011152
-25 0.420074349442379
-20 0.548327137546468
-15 0.657992565055762
-10 0.70817843866171
-5 0.777881040892193
0 0.803903345724907
5 0.823420074349442
10 0.820631970260223
15 0.820631970260223
20 0.819702602230483
25 0.819702602230483
30 0.819702602230483
};
\addlegendentry{SONNET}
\addplot [very thick, darkorange25512714]
table {%
-30 0.258364312267658
-25 0.258364312267658
-20 0.258364312267658
-15 0.258364312267658
-10 0.258364312267658
-5 0.258364312267658
0 0.258364312267658
5 0.258364312267658
10 0.258364312267658
15 0.258364312267658
20 0.258364312267658
25 0.258364312267658
30 0.258364312267658
};
\addlegendentry{GCC-PHAT}
\end{axis}

\end{tikzpicture}}
        %\includegraphics[width=\textwidth]{images/sim_noise_sensitivity_02t60.png}
        %\caption{Noise sensitivity evaluated \\ at $T_{60}$=0.2 s}
        \caption{}
        \label{fig:noise_sense}
    \end{subfigure}
    \begin{subfigure}{0.49\textwidth}
        \centering
        \scalebox{0.72}{% This file was created with tikzplotlib v0.10.1.
\begin{tikzpicture}

\definecolor{darkgray176}{RGB}{176,176,176}
\definecolor{darkorange25512714}{RGB}{255,127,14}
\definecolor{lightgray204}{RGB}{204,204,204}
\definecolor{steelblue31119180}{RGB}{31,119,180}

\begin{axis}[
legend cell align={left},
legend style={fill opacity=0.8, draw opacity=1, text opacity=1, draw=lightgray204},
tick align=outside,
tick pos=left,
x grid style={darkgray176},
xlabel={\(\displaystyle T_{60}\) (s)},
xmin=0, xmax=1,
xtick style={color=black},
y grid style={darkgray176},
ylabel={Inlier ratio @ 10cm},
ymin=0, ymax=1,
ytick style={color=black}
]
\addplot [very thick, steelblue31119180]
table {%
0.05 0.96392939370683
0.15 0.906876522390856
0.25 0.650780608052588
0.35 0.463863337713535
0.45 0.260786559755632
0.55 0.173116089613035
0.65 0.108839779005525
0.75 0.0870673952641166
0.85 0.0638578011849901
0.95 0.054726368159204
};
\addlegendentry{SONNET}
\addplot [very thick, darkorange25512714]
table {%
0.05 0.402148887183423
0.15 0.350196739741428
0.25 0.238564776773487
0.35 0.164695575996496
0.45 0.0988927071401298
0.55 0.0627970128988459
0.65 0.0403314917127072
0.75 0.0528233151183971
0.85 0.0302830809743252
0.95 0.0248756218905473
};
\addlegendentry{GCC-PHAT}
\end{axis}

\end{tikzpicture}}
        %\includegraphics[width=\textwidth]{images/sim_reverb_sensitivity_snr10.png}
        %\caption{Reverberation sensitivity evaluated at SNR=10dB}
        \caption{}
        \label{fig:reverb_sense}
    \end{subfigure}
    \caption{Results on the simulated data. \textbf{(a)} Noise sensitivity evaluated at $T_{60}$~=~0.2~s. Note that GCC-PHAT is very robust against white noise \textbf{(b)} Reverberation sensitivity evaluated at SNR~=~10~dB}
    \label{fig:sensitivity}
\end{figure}
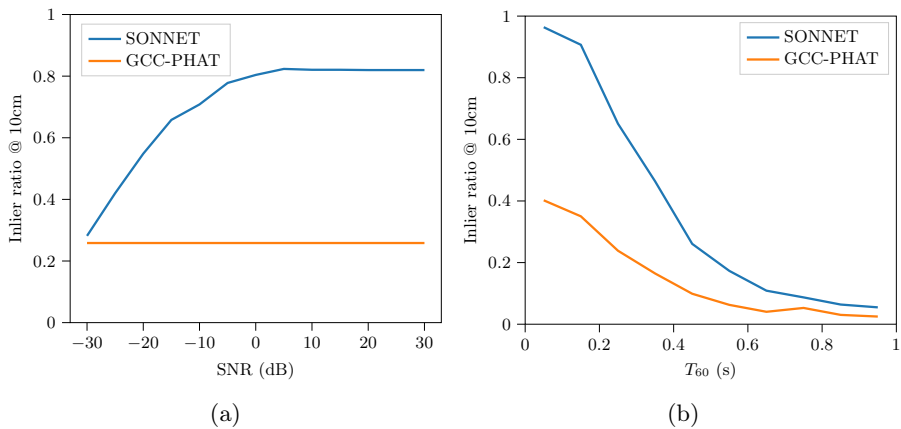

We have also evaluated our model's robustness to noise and reverberation using simulated data, see Fig.~\ref{fig:sensitivity}. We use accuracy at 10 cm as our main evaluation metric. To motivate this, we would like to highlight the distribution of the residuals. The residual distribution for all three detectors shown here seems to be well explained by a combination of a normal distribution (inliers) and a uniform distribution (outliers), a common combination of distributions in the area of robust estimation. Because of this, using mean squared error as an evaluation metric is both noisy and highly dependent on the space of values the model can estimate making comparisons between models more difficult. When using these detections for downstream tasks, because they contain outliers one probably need to use methods from robust estimation. Because of this, we think reporting inlier ratio, at a given inlier threshold, is a better evaluation metric.

The evaluation examples are created in the same manner as the training data, with the change that instead of using audio from \textit{Musan} we used the audio from \textit{tdoa\_20201016} (described in Section\ref{sec:real-data}) in the simulation. Our model outperforms GCC-PHAT in a wide range of reverberant and noise environments\change{, as shown in Fig~\ref{fig:sensitivity}}.

\subsection{Real Data}
\label{sec:real-data}
\begin{figure}
    \centering
    % This file was created with tikzplotlib v0.10.1.
\begin{tikzpicture}

\definecolor{darkgray176}{RGB}{176,176,176}
\definecolor{darkorange25512714}{RGB}{255,127,14}
\definecolor{lightgray204}{RGB}{204,204,204}
\definecolor{steelblue31119180}{RGB}{31,119,180}

\begin{axis}[
legend cell align={left},
legend style={
  fill opacity=0.8,
  draw opacity=1,
  text opacity=1,
  at={(0.97,0.03)},
  anchor=south east,
  draw=lightgray204
},
tick align=outside,
tick pos=left,
x grid style={darkgray176},
xlabel={Inlier threshold (m)},
xmin=0, xmax=0.5,
xtick style={color=black},
y grid style={darkgray176},
ylabel={Inlier ratio},
ymin=0, ymax=1,
ytick style={color=black}
]
\addplot [semithick, black, dashed, forget plot]
table {%
0.1 0
0.1 1
};
\addplot [very thick, steelblue31119180]
table {%
0 0
0.02 0.224415569559701
0.04 0.403441588153325
0.06 0.53614213514694
0.08 0.627914352862877
0.1 0.687836671450261
0.12 0.727558703021984
0.14 0.752750566751939
0.16 0.768778207607995
0.18 0.779398307023564
0.2 0.787634408602151
0.22 0.794102660094424
0.24 0.799645390070922
0.26 0.804681163037374
0.28 0.808929202803602
0.3 0.812810673654874
0.32 0.816328175370729
0.34 0.81965329340072
0.36 0.822806825981157
0.38 0.825793972671118
0.4 0.828705725754456
0.42 0.831617478837795
0.44 0.834599425968678
0.46 0.837466982799859
0.48 0.840716707223227
};
\addlegendentry{SONNET}
\addplot [very thick, darkorange25512714]
table {%
0 0
0.02 0.111070381231672
0.04 0.202286246126328
0.06 0.274830494374077
0.08 0.335441754539215
0.1 0.383490879973378
0.12 0.420610532226867
0.14 0.449954243880119
0.16 0.474293379921383
0.18 0.494483267818889
0.2 0.511293442316092
0.22 0.525376448077203
0.24 0.537423826459516
0.26 0.547718953432749
0.28 0.556358020839833
0.3 0.563949377092823
0.32 0.570698404775275
0.34 0.577252448992325
0.36 0.583413926499033
0.38 0.589229633327094
0.4 0.595151931116241
0.42 0.600928641251222
0.44 0.606658555354506
0.46 0.612625049395811
0.48 0.618638339468813
};
\addlegendentry{GCC-PHAT}
\end{axis}

\end{tikzpicture}
    \caption{Quantitative results on the dataset \textit{tdoa\_20201016} showing the probability of correct detection at different inlier thresholds. We have marked the 10 cm threshold which we use as our main evaluation metric.}
    \label{fig:acc_at}
\end{figure}
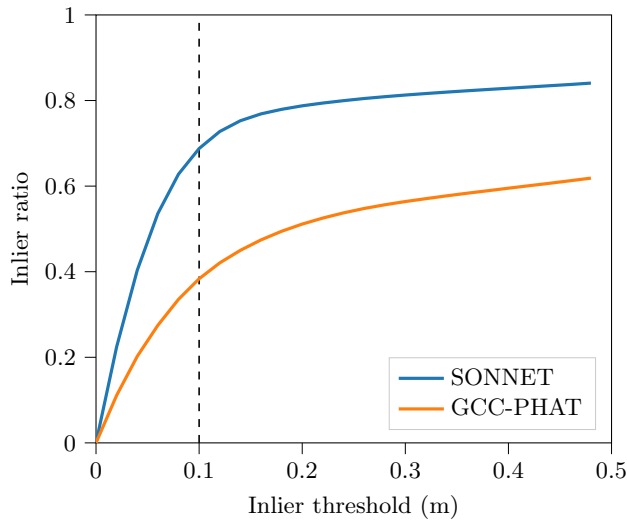

\begin{figure}
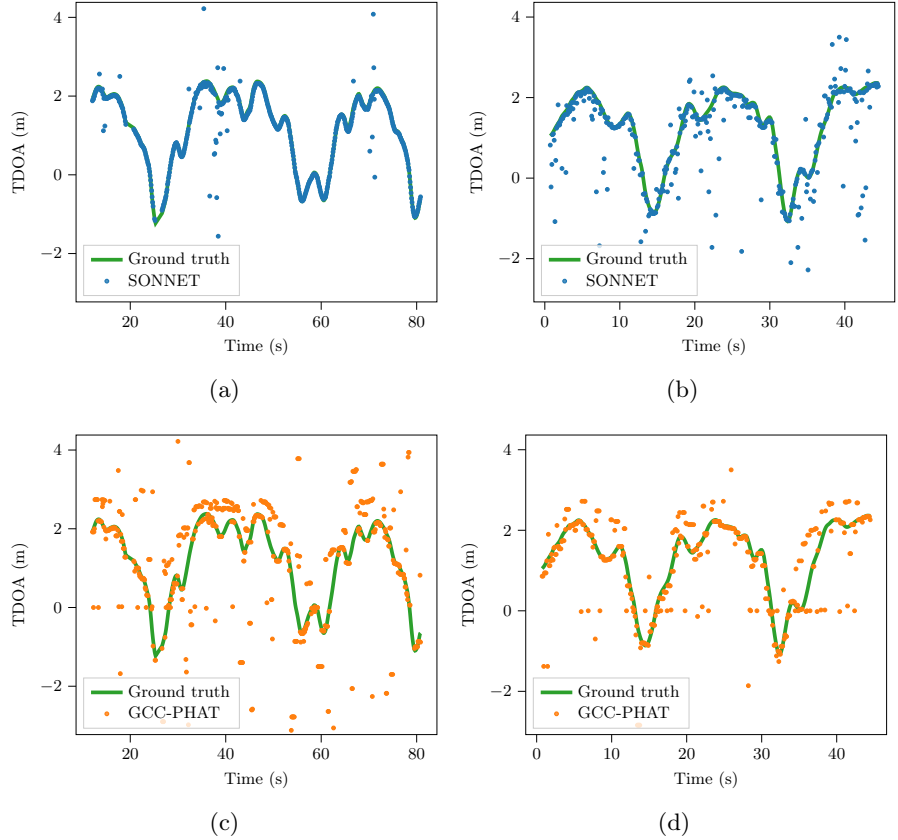

    \centering
    \begin{subfigure}{0.49\textwidth}
        \centering
        \scalebox{0.7}{\input{tikz-images/SONNET_on_music_0014}}
        \caption{}
    \end{subfigure}
    \begin{subfigure}{0.49\textwidth}
        \centering
        \scalebox{0.7}{\input{tikz-images/SONNET_on_chirp_0001}}
        \caption{}
    \end{subfigure}%
    \vspace{10pt}
    \hfill
    \begin{subfigure}{0.49\textwidth}
        \centering
        \scalebox{0.7}{\input{tikz-images/GCC-PHAT_on_music_0014}}
        \caption{}
    \end{subfigure}%
    \begin{subfigure}{0.49\textwidth}
        \centering
        \scalebox{0.7}{\input{tikz-images/GCC-PHAT_on_chirp_0001}}
        \caption{}
    \end{subfigure}%
    ~
    \caption{Qualitative results of the estimated TDOA values on the dataset \textit{tdoa\_20201016}. \textbf{(a)} and \textbf{(c)} correspond to the recording \textit{music\_0014} while \textbf{(b)} and \textbf{(d)} correspond to \textit{chirp\_0001}. The microphone paired used for all four plots are microphone 1 and microphone 6. SONNET significantly outperforms GCC-PHAT when music is played while also achieving a performance gain  when chirp sounds are played.}
    \label{fig:qual_res}
\end{figure}

Arguably, the most important evaluation of our model is the results on real world data. We have evaluated our model on the \textit{tdoa\_20201016} dataset provided by \cite{astrom2021extenstion}. The advantage of using this dataset is that it contains ground truth values for the TDOA for any pair of two microphones. The dataset also contains recordings without accompanying ground truth but these were not used in our evaluation. The dataset is recorded in 96 kHz, which means that we have to down-sample, since our model is trained on 16 kHz. The total playing time of the speaker over all the experiments is around 600 s with 12 microphones recording. Using a window overlap of 5/6, we get 384648 pairs of windows to estimate TDOA on. As shown in Fig~\ref{fig:acc_at}, the learned models significantly outperforms GCC-PHAT. We also show qualitative results over some of the recordings in the dataset in Fig~\ref{fig:qual_res}.  

\subsection{Downstream Application}
Since the main reason for studying TDE is its use in downstream applications, we have evaluated our models on the task of self-calibration using the \cite{astrom2021extenstion} dataset. In self-calibration the goal is to estimate the 3D geometry of both the receivers and senders using only the TDOA values as input, i.e. no prior position information. 

To do this, we used the TDOA values acquired as input to a published self-calibration system \cite{larsson2021fast,astrom2021extenstion}. We then compare our estimated 3D positions with the ground truth positions provided in the dataset. In order to be able to do this comparison, we need to fix the gauge freedom in the solution, in this case the solution and ground truth might differ by a Euclidean transformation. We estimate this Euclidean transformation using the receiver positions of our solution compared to the ground truth receiver positions. This is similar to how to evaluate maps found using Structure-from-Motion (SfM) or Simultaneous Localization And Mapping (SLAM). After applying the Euclidean transformation to the solution, the residuals are then computed as the distances between corresponding receivers in the solution and ground truth. 

\begin{table}[]
    \centering
    \caption{RMS error of the receivers for the estimated 3D geometry after registration to the ground truth. Estimation is done using TDOA values from SONNET or GCC-PHAT. Experiments missing a value have an error larger than 1 m.}
    %Note, the maximum rms error shown is 1 m, so for the experiments missing a value we have error larger than 1 m which we consider as not having found a correct solution. As we can see the found 3D reconstructions are significantly better on both chirp and music data using SONNET for TDE instead of GCC-PHAT.}
    \vspace{10pt}
    \label{tab:downstream-rms}
    \begin{tabular}{l@{\hskip 10pt}c@{\hskip 6pt}c}
    \toprule
%        \hline
         Experiment  & SONNET (m) & GCC-PHAT (m)\\
         \midrule
    %    \hline
chirp\_0001 & \textbf{0.05} & 0.80 \\
chirp\_0002 & \textbf{0.05} & 0.17 \\
chirp\_0004 & \textbf{0.05} & 0.40 \\
iregchirp\_0006 & \textbf{0.06} & 0.54 \\
iregchirp\_0007 & \textbf{0.04} & 0.59 \\
music\_0008 & \textbf{0.07} & -- \\
music\_0009 & \textbf{0.06} & -- \\
music\_0010 & \textbf{0.04} & 0.31 \\
music\_0011 & \textbf{0.05} & -- \\
music\_0012 & \textbf{0.04} & -- \\
music\_0013 & \textbf{0.03} & 0.35 \\
music\_0014 & \textbf{0.04} & 0.28 \\
music\_0015 & \textbf{0.04} & 0.10 \\
metronom\_0021 & \textbf{0.16} & -- \\
metronom\_0022 & \textbf{0.10} & -- \\
\midrule
median & \textbf{0.05} & 0.59 \\
        \bottomrule
    \end{tabular}
\end{table}

\begin{figure}
    \centering
    %\scalebox{0.7}{\input{tikz-images/3d_plot.tex}}
    \includegraphics[width=0.8\textwidth]{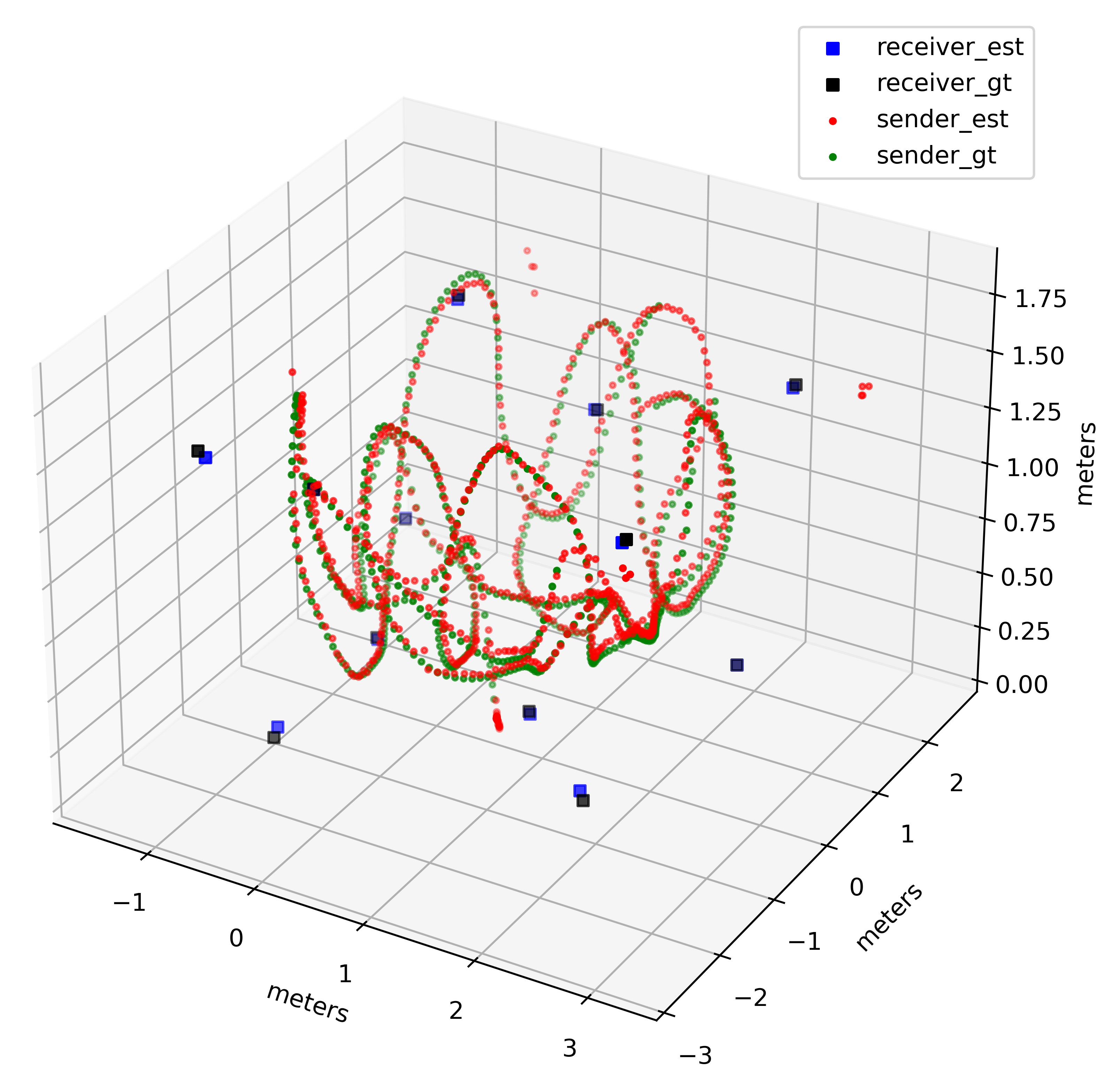}
    \caption{Example of 3D reconstruction, on the experiment \textit{music\_0014}}
    \label{fig:3d-reconstruct}
\end{figure}

As can be seen in Table \ref{tab:downstream-rms}, using the TDOA values from our learned models makes the self-calibration system converge to good solutions on all of the experiments. This is a significant improvement compared to using GCC-PHAT for which the system only manages to converge on some of the experiments and even when it converges it has larger errors. An example of a 3D reconstruction resulting from using the learned model together with the self-calibration system can be seen in Fig \ref{fig:3d-reconstruct}.

% \begin{figure}
%     \centering
%     \begin{subfigure}{0.49\textwidth}   
%         \centering
%         \includegraphics[width=\textwidth]{images/system_result_sonnet.png}
%         %\caption{Noise sensitivity evaluated \\ at $T_{60}$=0.2 s}
%         \caption{}
%         \label{fig:downstream-rms-sonnet}
%     \end{subfigure}
%     \begin{subfigure}{0.49\textwidth}
%         \centering
%         \includegraphics[width=\textwidth]{images/system_result_gcc.png}
%         %\caption{Reverberation sensitivity evaluated at SNR=10dB}
%         \caption{}
%         \label{fig:downstream-rms-gcc}
%     \end{subfigure}
%     \caption{RMS error of the estimated 3D geometry using the TDOA values estimated by \textbf{(a)} SONNET \textbf{(b)} GCC-PHAT. Note, the maximum rms error shown is 1 m, so for the experiments missing a marker we have error larger than 1 m which we consider as not having found a correct solution. As we can see the found 3D reconstructions are significantly better on both chirp and music data using SONNET for TDE instead of GCC-PHAT.}
%     \label{fig:downstream-rms}
% \end{figure}
\vspace{-3pt}

\subsection{Ablation studies}
Earlier works which use simulated data to train models to solve TDE have not been demonstrated to generalize to real world data. We claim that we can achieve a good generalization to novel real world data by: scaling the dataset, simulating sound source with movement and directionality. To show that these three changes are helpful we have performed two ablation studies.

\begin{figure}
    \centering
    \begin{subfigure}{0.49\textwidth}
        \centering
        \scalebox{0.7}{% This file was created with tikzplotlib v0.10.1.
\begin{tikzpicture}

\definecolor{crimson2143940}{RGB}{214,39,40}
\definecolor{darkgray176}{RGB}{176,176,176}
\definecolor{darkorange25512714}{RGB}{255,127,14}
\definecolor{forestgreen4416044}{RGB}{44,160,44}
\definecolor{lightgray204}{RGB}{204,204,204}
\definecolor{steelblue31119180}{RGB}{31,119,180}

\begin{axis}[
legend cell align={left},
legend style={
  fill opacity=0.8,
  draw opacity=1,
  text opacity=1,
  at={(0.97,0.03)},
  anchor=south east,
  draw=lightgray204
},
tick align=outside,
tick pos=left,
x grid style={darkgray176},
xlabel={Inlier threshold (m)},
xmin=0, xmax=0.5,
xtick style={color=black},
y grid style={darkgray176},
ylabel={Inlier ratio},
ymin=0, ymax=1,
ytick style={color=black}
]
\addplot [semithick, black, dashed, forget plot]
table {%
0.1 0
0.1 1
};
%\addplot [very thick, steelblue31119180]
\addplot [very thick, sonnet_red]
table {%
0 0
0.02 0.17070412429026
0.04 0.314191156589921
0.06 0.423324702065265
0.08 0.501617062873068
0.1 0.554525176265053
0.12 0.590820698404775
0.14 0.615066242382646
0.16 0.63257055801668
0.18 0.64494290884133
0.2 0.654250119589859
0.22 0.66157889810944
0.24 0.667475198103201
0.26 0.672147001934236
0.28 0.675651504752397
0.3 0.67877383997837
0.32 0.68124363054013
0.34 0.683443044029866
0.36 0.685554065015287
0.38 0.687623489528088
0.4 0.68949273101641
0.42 0.691510159938437
0.44 0.693613381585242
0.46 0.695604812711882
0.48 0.697684636342838
};
\addlegendentry{SONNET}
%\addplot [very thick, darkorange25512714]
\addplot [very thick, sonnet_magenta]
table {%
0 0
0.02 0.175615107838855
0.04 0.323610157858614
0.06 0.438356627358
0.08 0.521544373037166
0.1 0.578874711424471
0.12 0.619137497140242
0.14 0.647202117260456
0.16 0.667210020590254
0.18 0.681937771676962
0.2 0.693163619725047
0.22 0.701649300139348
0.24 0.708343732451488
0.26 0.713847465734906
0.28 0.718147501091907
0.3 0.721652003910068
0.32 0.724506561843556
0.34 0.727129739398099
0.36 0.729407146273996
0.38 0.731671554252199
0.4 0.733790374576236
0.42 0.735849399970883
0.44 0.73804621368108
0.46 0.740193631580042
0.48 0.742421642644704
};
\addlegendentry{SONNET (d)}
%\addplot [very thick, forestgreen4416044]
\addplot [very thick, sonnet_aqua]
table {%
0 0
0.02 0.217895322476654
0.04 0.393068467793931
0.06 0.524180549489403
0.08 0.613761153054221
0.1 0.670758719660573
0.12 0.708325533994717
0.14 0.732282502443793
0.16 0.746604687922464
0.18 0.755761111457748
0.2 0.762642727896674
0.22 0.768058068675776
0.24 0.772534889041409
0.26 0.776333166947443
0.28 0.779536095339115
0.3 0.78233605790229
0.32 0.784850044716208
0.34 0.787192446080572
0.36 0.789589442815249
0.38 0.791851251013914
0.4 0.794053264283189
0.42 0.796304673363699
0.44 0.798803061500385
0.46 0.801491233543395
0.48 0.804267798090722
};
\addlegendentry{SONNET (m)}
%\addplot [very thick, crimson2143940]
\addplot [very thick, sonnet]
table {%
0 0
0.02 0.224415569559701
0.04 0.403441588153325
0.06 0.53614213514694
0.08 0.627914352862877
0.1 0.687836671450261
0.12 0.727558703021984
0.14 0.752750566751939
0.16 0.768778207607995
0.18 0.779398307023564
0.2 0.787634408602151
0.22 0.794102660094424
0.24 0.799645390070922
0.26 0.804681163037374
0.28 0.808929202803602
0.3 0.812810673654874
0.32 0.816328175370729
0.34 0.81965329340072
0.36 0.822806825981157
0.38 0.825793972671118
0.4 0.828705725754456
0.42 0.831617478837795
0.44 0.834599425968678
0.46 0.837466982799859
0.48 0.840716707223227
};
\addlegendentry{SONNET (d+m)}
\end{axis}

\end{tikzpicture}}
        \caption{}
        \label{fig:ablation_aug}
        \end{subfigure}
    \begin{subfigure}{0.49\textwidth}
        \centering
        \scalebox{0.7}{% This file was created with tikzplotlib v0.10.1.
\begin{tikzpicture}

\definecolor{darkgray176}{RGB}{176,176,176}
\definecolor{steelblue31119180}{RGB}{31,119,180}

\begin{axis}[
log basis x={10},
tick align=outside,
tick pos=left,
x grid style={darkgray176},
xlabel={Relative dataset size},
xmin=0.01, xmax=4,
xmode=log,
xtick style={color=black},
y grid style={darkgray176},
ylabel={Inlier ratio @ 10 cm},
ymin=0, ymax=1,
ytick style={color=black}
]
\addplot [semithick, steelblue31119180, mark=*, mark size=3, mark options={solid}, only marks]
table {%
2 0.722634720575695
1 0.687836671450261
0.5 0.628722884299411
0.3 0.581872257232587
0.2 0.543983070235644
0.15 0.507326178740043
0.1 0.420529939061168
0.05 0.300284935837441
0.02 0.0486236767122148
};
\end{axis}

\end{tikzpicture}}
        \caption{}
    \label{fig:ablation_dataset}
    \end{subfigure}%
    \hfill
    ~
    \caption{Results from the ablation studies. \textbf{(a)} Ablation study on the effect of introducing the simulation augmentations: sound source movement (m) and directionality (d). Introducing sound source movement gives a larger performance gain. \textbf{(b)} Ablation study on the effect of the size of the simulated training dataset, when model is evaluated on the real data from section \ref{sec:real-data}. The size of the dataset is given in relative sizes to the dataset SONNET was trained on.}
    \label{fig:both_abl}
\end{figure}
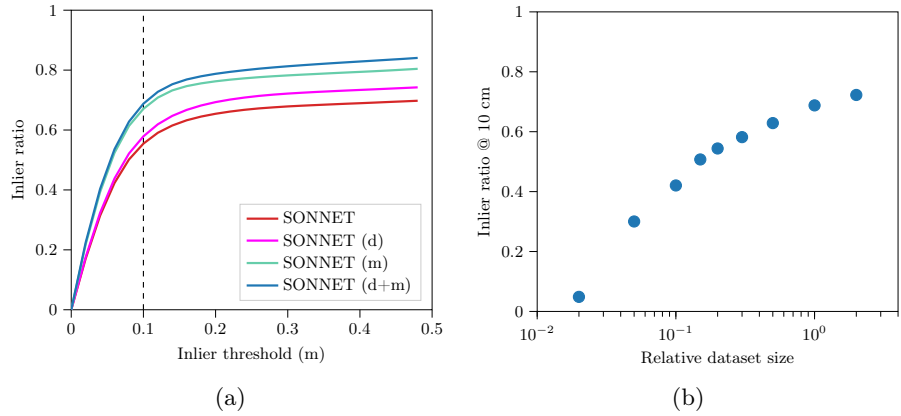

In the first ablation study we changed how the dataset was generated by including or excluding the simulation augmentations: moving the sound source or directionality of the sound source. For each of the four configuration we trained a separate model in the same way as the full SONNET model. Each of the models were then evaluated on \textit{tdoa\_20201016} in the same way as in section \ref{sec:real-data}, and the results are shown in Fig \ref{fig:ablation_aug}. As we can see, using learning based methods on stationary omnidirectional data outperforms GCC-PHAT. However, we can further improve the method by augmenting the simulation.

% \begin{figure}
%     \centering
%     \scalebox{0.7}{\input{tikz-images/ablation_study_sim.tex}}
%     %\includegraphics[width=0.49\textwidth]{images/ablation_study_plot.png}
%     \caption{Ablation study on the effect of introducing the simulation augmentations: sound source movement (m) and directionality (d). Introducing sound source movement gives a larger performance gain.}
%     \label{fig:ablation_aug}
% \end{figure}

For the second ablation study, we trained models on different sizes of the training dataset. The models were then evaluated on \textit{tdoa\_20201016} in the same way as in section \ref{sec:real-data}, the results are shown in Fig \ref{fig:ablation_dataset}. As we can see, having a large enough dataset is important for generalization and scaling up the dataset might be a way to improve the model further.

% \begin{figure}
%     \centering
%     \scalebox{0.7}{\input{tikz-images/ablation_study_size}}
%     %\includegraphics[width=0.45\textwidth]{images/ablation_dataset_size_linear.png}
%     %\includegraphics[width=0.45\textwidth]{images/ablation_dataset_size_log.png}
%     \caption{Ablation study on the effect of the size of the simulated training dataset, when model is evaluated on the real data from section \ref{sec:real-data}. The size of the dataset is given in relative sizes to the dataset SONNET was trained on.}
%     \label{fig:ablation_dataset}
% \end{figure}

% \begin{figure}
%     \centering
%     \includegraphics[width=0.8\textwidth]{images/ablation_model_size.png}
%     \caption{Ablation study on the effect of increasing or decreasing the size of the model, evaluated on the real data from section \ref{sec:real-data} \todo{(I think we should remove this) Scaling up and down the network in a unique way is difficult. The only thing I can show with this plot is that the result is not sensitive to model size. If you wanted to deploy a small model you would not train a small model but rather train a big model and distill a small model, so there is not an argument for appealing to people who want a very samll network. }}
%     \todo{? om inte beror på modellstorlek är det väl bättre att bara använda noll parametrar? (någonstans tappar man rimligen prestanda?), beroendet är väl även i samma storleksordning som ablation i fig 8?}
%     \label{fig:ablation_dataset}
% \end{figure}

\label{exp:abl}

%Conclusions
\section{Conclusions}
As we have demonstrated in this paper, combining the ability to simulate data with learning based methods is a promising direction for further studies. In this paper we have shown that it is possible to improve on TDE, a key task when performing audio based localization. However, using simulation together with learning based methods is also a ripe area for further studies, since it enables an approach to harder versions of the TDE problem. Such examples include using multiple sound sources, multipath components of the sound, or utilizing the information from more than two microphones at the same time. Since we can simulate such data with ground truth, it might be possible to create detectors for such problems. It is our belief that this paper is a key stepping stone for further studies in the area.

% Refrences -----------------------------------------------

\bibliographystyle{splncs04}
\bibliography{ref}

\end{document}